\documentclass[aps,pra,a4paper,twocolumn,noshowpacs,groupedaddress,floatfix,superscriptaddress,amsmath,amssymb,citeautoscript]{revtex4-2}

\pdfoutput=1

\usepackage[T1]{fontenc}
\usepackage[latin1]{inputenc}
\usepackage{dcolumn, bm, graphicx, color, booktabs, microtype, afterpage}
\graphicspath{{Figures/}}
\usepackage[charter,greekuppercase=italicized]{mathdesign}
\usepackage{sidecap}
\renewcommand{\figurename}{\bf\textsf{Figure}\sffamily}
\makeatletter\renewcommand{\fnum@figure}[1]{\textbf{\sffamily\figurename~\thefigure~|\,}}\makeatother

\newcommand{\sfo}{Sr$_3$Fe$_2$O$_7$}

\definecolor{NatureBlue}{rgb}{0.012,0.3,0.63}

\usepackage[colorlinks,plainpages=false,linkcolor=NatureBlue,urlcolor=NatureBlue,citecolor=NatureBlue,pdfpagemode=UseNone,pdfstartview=FitBH]{hyperref}

\begin{document}

\title{\fontsize{22pt}{26pt}\selectfont\sffamily\bfseries\textcolor{NatureBlue}{Reentrant multiple-$\mathbf{q}$ magnetic order and a ``spin-cholesteric'' phase in Sr$_3$Fe$_2$O$_7$}}

\author{\large\sffamily N.\,D.\,Andriushin}
\affiliation{\sffamily Institut f\"ur Festk\"orper- und Materialphysik, Technische Universit\"at Dresden, 01069 Dresden, Germany}
\author{\large\sffamily J.\,Muller}
\affiliation{\sffamily Zernike Institute for Advanced Materials, University of Groningen, 9747 AG Groningen, The Netherlands}
\author{\large\sffamily N.\,S.\,Pavlovskii}
\affiliation{\sffamily Institut f\"ur Festk\"orper- und Materialphysik, Technische Universit\"at Dresden, 01069 Dresden, Germany}
\author{\large\sffamily J.\,Grumbach}
\affiliation{\sffamily Institut f\"ur Festk\"orper- und Materialphysik, Technische Universit\"at Dresden, 01069 Dresden, Germany}
\author{\large\sffamily S.\,Granovsky}
\affiliation{\sffamily Institut f\"ur Festk\"orper- und Materialphysik, Technische Universit\"at Dresden, 01069 Dresden, Germany}
\author{\large\sffamily Y.\,V.\,Tymoshenko}
\affiliation{\sffamily Institut f\"ur Festk\"orper- und Materialphysik, Technische Universit\"at Dresden, 01069 Dresden, Germany}
\author{\large\sffamily O.\,Zaharko}
\affiliation{\sffamily Laboratory for Neutron Scattering and Imaging, Paul Scherrer Institute, CH-5232 Villigen, Switzerland\looseness=-1}
\author{\large\sffamily A.\,Ivanov}
\affiliation{\sffamily Institut Laue-Langevin, 71 avenue des Martyrs, CS 20156, 38042 Grenoble Cedex 9, France}
\author{\large\sffamily J.\,Ollivier}
\affiliation{\sffamily Institut Laue-Langevin, 71 avenue des Martyrs, CS 20156, 38042 Grenoble Cedex 9, France}
\author{\large\sffamily M.\,Doerr}
\affiliation{\sffamily Institut f\"ur Festk\"orper- und Materialphysik, Technische Universit\"at Dresden, 01069 Dresden, Germany}
\author{\large\sffamily B.\,Keimer}
\affiliation{\sffamily Max-Planck-Institut f\"ur Festk\"orperforschung, Heisenbergstra{\ss}e 1, 70569 Stuttgart, Germany}
\author{\large\sffamily M.\,Mostovoy}
\affiliation{\sffamily Zernike Institute for Advanced Materials, University of Groningen, 9747 AG Groningen, The Netherlands}
\author{\large\sffamily D.\,S.\,Inosov}
\email{dmytro.inosov@tu-dresden.de}
\affiliation{\sffamily Institut f\"ur Festk\"orper- und Materialphysik, Technische Universit\"at Dresden, 01069 Dresden, Germany}
\affiliation{\sffamily W\"urzburg-Dresden Cluster of Excellence on Complexity and Topology in Quantum Matter\,---\,\textit{ct.qmat}, TU Dresden}
\author{\large\sffamily D.\,C.\,Peets}
\affiliation{\sffamily Institut f\"ur Festk\"orper- und Materialphysik, Technische Universit\"at Dresden, 01069 Dresden, Germany}
\affiliation{\sffamily Max-Planck-Institut f\"ur Festk\"orperforschung, Heisenbergstra{\ss}e 1, 70569 Stuttgart, Germany}

\begin{abstract}
\fontsize{9pt}{11pt}\noindent\textbf{Spin-nematic and spin-smectic phases have been reported in magnetic materials, which break rotational symmetry while preserving translational symmetry along certain directions. However, until now the analogy to liquid crystals remained incomplete because no magnetic analog of cholesteric order was known. Here we show that the bilayer perovskite Sr$_3$Fe$_2$O$_7$, previously believed to adopt a simple single-$\mathbf{q}$ spin-helical order, hosts two distinct types of multi-$\mathbf{q}$ spin textures and the first ``spin-cholesteric''. Its ground state represents a novel multi-$\mathbf{q}$ spin texture with unequally intense spin modulations at the two ordering vectors. This is followed in temperature by the new ``spin-cholesteric'' phase with spontaneously broken chiral symmetry, in which the translational symmetry is broken only along one of the crystal directions while the weaker orthogonal modulation melts, giving rise to intense short-range dynamical fluctuations. Shortly before the transition to the paramagnetic state, vortex-crystal order spanned by two equivalent $\mathbf{q}$ vectors emerges. The ``spin-cholesteric'' phase completes the spin analogy with liquid crystals and renders Sr$_3$Fe$_2$O$_7$ a touchstone for studying transitions among multiple-$\mathbf{q}$ spin textures in a centrosymmetric host.}
\end{abstract}

\citestyle{nature}
\maketitle

Liquid crystals, which can form smectic, nematic, and cholesteric phases depending on their symmetry, take an intermediate position between crystals and liquids~\cite{chaikin1995principles}. Magnetic phases with similar symmetry properties arise upon ordering of spins in a magnet, where conventional magnetic order with broken translational symmetry is analogous to crystalline order, while paramagnets and spin-liquid states mirror the characteristics of liquids. The liquid-crystal states exhibit combinations of broken rotational and nonbroken or only partially broken translational symmetries\,---\,a trait that finds analogies in electronic and spin systems: electronic liquid crystals~\cite{Kivelson_1998}, spin nematics~\cite{Hinkov_2008, Chuang_2010} and spin smectics~\cite{Lebert_2019, Cruddas_2021}. The intricate interplay of symmetry rules, degeneracy, and strong fluctuations makes these states particularly fascinating. However, the magnetic analogue of chiral nematic liquid crystal, namely a cholesteric, has not yet been reported.

\begin{figure}
  \includegraphics[width=\columnwidth]{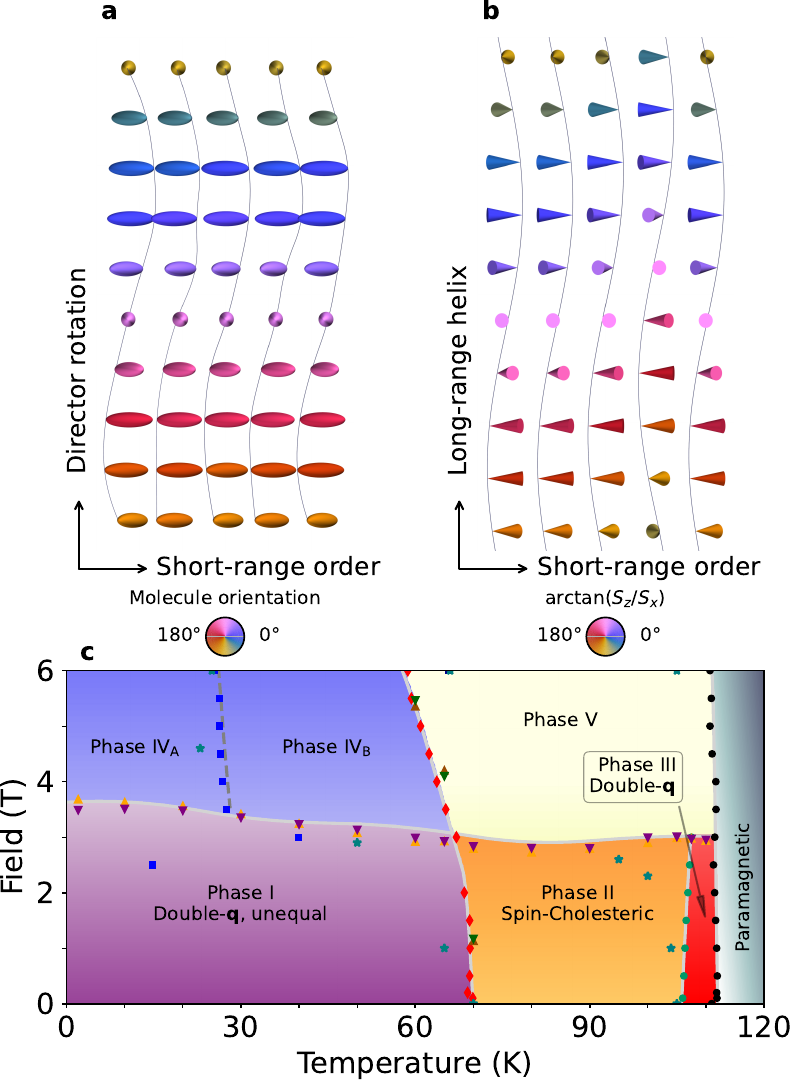}
	\caption{(a), Schematic drawing of molecular ordering in a cholesteric liquid crystal. (b), Schematic diagram of spin ordering in a ``spin-cholesteric'' state. (c), Field-temperature magnetic phase diagram for \sfo, $\mathbf{H}\parallel [110]$, based on magnetization and dilatometry measurements. Triangles denote transitions extracted from $M$($H$) data; squares, diamonds and circles indicate $M$($T$) points; and stars denote dilatometry transitions \cite{PhaseDiag}.}
	\label{PD}
\end{figure}

The ``spin-cholesteric'' state in \sfo\ is the missing link in this sequence of electronic liquid crystals. The ground-state magnetic structure of \sfo, the Ruddlesden-Popper bilayer analogue of cubic perovskite SrFeO$_3$~\cite{Ishiwata_2020}, was previously refined from neutron powder diffraction as a single-$\mathbf{q}$ helix with incommensurate propagation vector $\mathbf{Q} = (\xi\,\xi\,1)$~\cite{Peets2014}. However, these experiments could not distinguish a coexistence of multiple magnetic domains from multi-$\mathbf{q}$ order. Very recently, we found a vastly more complex magnetic phase diagram in \sfo\ [Fig.~\ref{PD}(c)], suggestive of multi-$\mathbf{q}$ order in at least one phase~\cite{PhaseDiag}. Here we identify the magnetic order in \sfo\ in a single-domain state prepared by domain selection in an external magnetic field. By means of single-crystal neutron diffraction and spectroscopic methods, we show that \sfo\ is a unique compound hosting two distinct types of double-$\mathbf{q}$ magnetic order divided by an exotic fluctuation-driven state. This intermediate phase is the long-sought ``spin-cholesteric'' state characterized by a fusion of short-range and long-range spin orders and symmetry-breaking rules analogous to those found in chiral nematic liquid crystals. 

The molecules in a nematic liquid crystal are oriented toward the director axis, which finds its counterpart in the spin direction in case of magnetic analogue. Spin-nematic order lacks rotational symmetry due to the presence of a preferred spin orientation and simultaneously does not break time reversal invariance thanks to fluctuations or frustration~\cite{Podolsky_2005}. In a cholesteric liquid crystal, chiral symmetry is additionally broken due to a systematic rotation of the director axis among the layers [Fig.~\ref{PD}(a)]. Hence, the ``spin-cholesteric'' state [Fig.~\ref{PD}(b)] is expected to have broken chiral and time-reversal symmetry, while translational symmetry is broken along one direction.  In the other direction only short-range correlations should be preserved. Our neutron scattering data evidence that these conditions are fulfilled in the \sfo\ mid-temperature phase.

\begin{figure*}
\center{\includegraphics[width=0.99\linewidth]{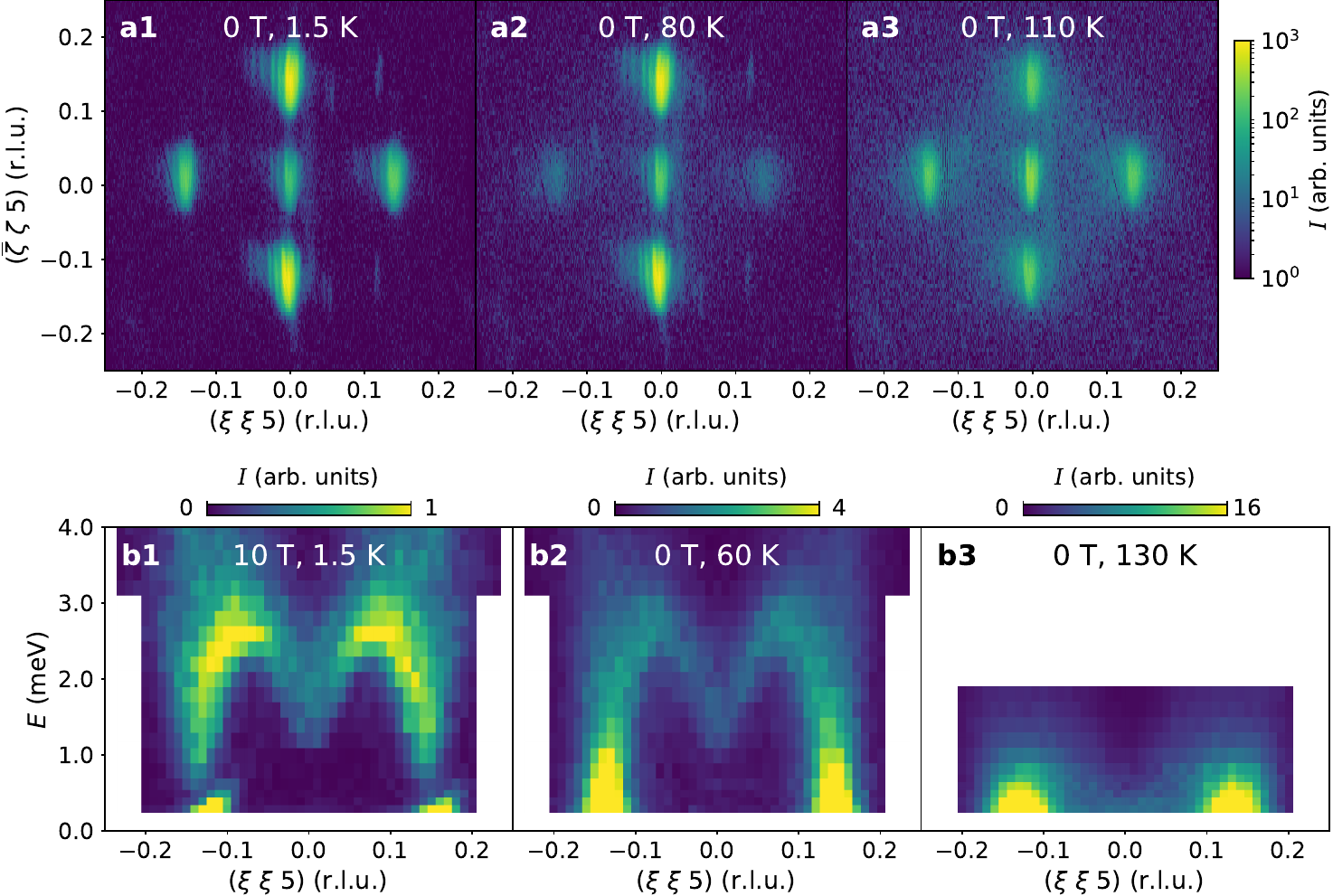}}
	\caption{Neutron scattering measurements on \sfo. (a1--a3), ZEBRA diffraction scans capturing in-plane and out-of-plane magnetic peaks simultaneously; the central peak at $\mathbf{Q} = (0~0~5)$ is a nonmagnetic nuclear reflection. Peaks are elongated along [$\overline{1}10$] due to vertical focusing. (b1--b3), Inelastic scattering intensity for the suppressed magnetic peaks (THALES triple-axis spectrometer).}
	\label{CMP}
\end{figure*}

\begin{figure*}
	\center{\includegraphics[width=1\linewidth]{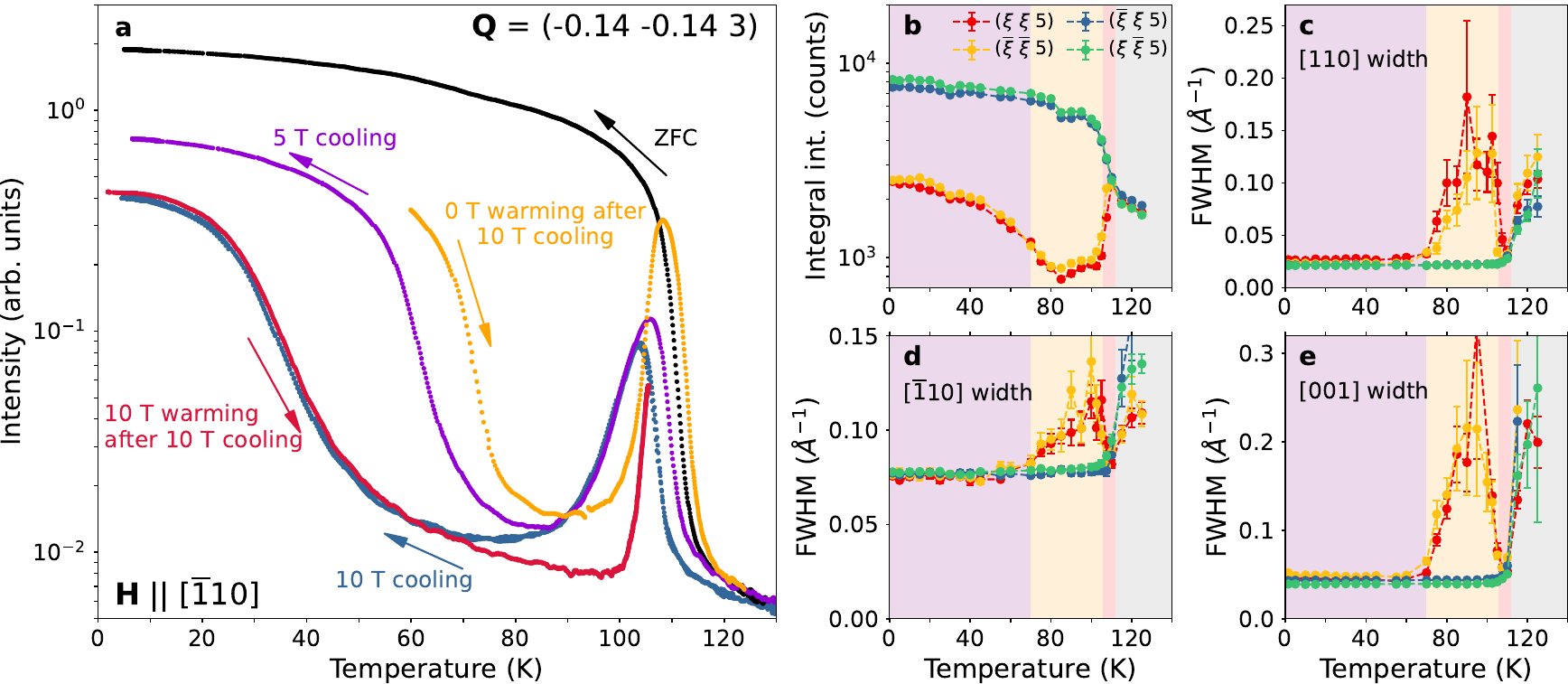}}
	\caption{Neutron scattering on \sfo\ in an external magnetic field $\mathbf{H} \parallel [\overline{1}\,1\,0]$. (a) Temperature dependence of elastic scattering intensity of magnetic peak $\mathbf{Q} = (-0.14, -0.14, 3)$ from triple-axis spectrometer THALES. (b), (c), (d), (e) Zero-field-warming data from single-crystal diffraction (ZEBRA) after cooling in field: integral intensity and peak widths along three orthogonal directions.}
	\label{I_vs_T}
\end{figure*}

The magnetic phase diagram of \sfo, partially reproduced for $\mathbf{H} \parallel [\overline{1}\,1\,0]$ in Fig.~\ref{PD}(c), includes three magnetic phases at low field, two of which exhibit field training indicative of the preparation of single-domain states if cooled in a field of $\gtrsim$~3~T (details on magnetization data can be found in Supplementary Sec.~\ref{sec:SMmagnetiz} and Ref.~\citenum{PhaseDiag}). For characterization of the low-field phases, our \sfo\ sample was prepared in a single-domain magnetic state by cooling to 1.5~K in a field of 5~T, after which the field was removed and high-statistics neutron diffraction maps in the ($HK5$) plane were measured on warming Fig.~\ref{CMP}(a). In tetragonal symmetry, the four magnetic satellites $\mathbf{Q} = (\pm\xi\mp\!\xi\,5)$ comprise two groups parallel and perpendicular to an external field $\mathbf{H} \parallel [\overline{1}\,1\,0]$, as the helical order is suppressed prefentially by fields $\mathbf{H}\perp\mathbf{Q}$. We indeed measure unequal intensities in two of the three phases: the peaks aligned with the field are favored, and the orthogonal ones are suppressed. The full temperature dependencies of the widths and integral intensities of all four peaks are shown in Fig.~\ref{I_vs_T}(b--e). The magnetization and neutron diffraction data indicate that domain selection is essential for detecting the transition from phase II to phase III.

In the low-temperature phase~I [Fig.~\ref{CMP}(a1)], two orthogonal modulations with unequal amplitudes form a nontrivial ordered state. The difference in intensity is remarkable, because in conventional skyrmion- and hedgehog-lattice phases all Fourier components of the spin modulation are equally developed. In \sfo, the suppressed and favored peaks differ by a factor of $\sim$3 in integral intensity, indicating an unequal double-$\mathbf{q}$ magnetic state intermediate between a spin helix and skyrmion lattice (a similar state was discussed theoretically in another system~\cite{Ozawa2016}). Non-zero intensity in the suppressed peaks cannot be attributed to incomplete domain selection of a single-$\mathbf{q}$ helicoidal structure\,---\,in such a scenario, warming would restore intensity in the suppressed domains as the system thermalized. We observe the opposite. At $T \approx 70$~K the suppressed peak intensity spontaneously {\itshape collapses} [Fig.~\ref{I_vs_T}(a)] upon entering phase II, indicating that the domain selection is retained, and these are distinct, well-defined phases.

In phase~II, the suppressed peaks lose intensity dramatically and become diffuse [Fig.~\ref{CMP}(a2)], while the favored peaks remain unchanged. Long-range order is preserved for one propagation vector, while the orthogonal reflections evidence only short-range correlations [Fig.~\ref{I_vs_T}(b--e)], in a novel failed multi-$\mathbf{q}$ phase. The \sfo\ magnetism in phase II provides the closest analogy to a ``spin-cholesteric'' state, i.e. a chiral variant of a spin nematic. The long-range favored $\mathbf{q}$-vector entails a spin helix (as is apparent from domain selection), introducing chiral- and time-reversal-symmetry breaking. The translational symmetry is only partially broken: the diffuse suppressed peaks provide clear evidence that spins are correlated in this direction only on short lengthscales, similarly to molecules in a liquid. These symmetry-breaking rules match those in cholesteric liquid crystals~\cite{Mitov_2012}. Another variant of a chiral spin nematic state was discussed in Ref.~\citenum{Onoda_2007}, however, unlike \sfo\ it preserves translation symmetry completely. 

The phase transition around 110~K from the ``spin-cholesteric'' phase into phase III, recently suggested as a candidate multi-$\mathbf{q}$ phase based on magnetization data~\cite{PhaseDiag}, is associated with redistribution of intensity between favored and suppressed peaks. The diffuse peaks spontaneously recondense above the transition, rendering all four magnetic satellites equally intense and resolution limited without any external influence besides temperature. Therefore, phase III in \sfo\ corresponds to reentrant multi-$\mathbf{q}$ order spanned by two orthogonal spin modulations. If this were a conventional skyrmion phase, this would imply heating into a lower-entropy state, so this phase is presumably also exotic.

Inconsistencies in the transition between phases II and III in Fig.~\ref{I_vs_T}(a) are likely associated with relaxation of magnetic domains, which depends on field conditions. The order parameter relaxation at 100~K can be described with a stretched exponential~\cite{Johnston_2006} with a stretching parameter $\beta =$~0.5, and has a characteristic timescale of $\tau =$~4400~s, comparable to the timescale of a diffraction measurement (for details see Supplementary Sec.~\ref{sec:SMrelax}).

Magnetic excitations and quasi-elastic scattering in \sfo\ were investigated using neutron spectroscopy. At 1.5~K [Fig.~\ref{CMP}(b1)], strong magnetic excitations feature broad branches above the suppressed magnetic satellites, as well as quasi-elastic intensity from the suppressed magnetic peaks. The map was collected in a field of 10~T; however, excitations in zero field are similar, differing only by a 0.1--0.2~meV shift (neutron scattering in field is discussed in Supplementary Sec.~\ref{sec:SMfield}). Phase I has two types of ordering vectors, whose excitations are different as confirmed using time-of-flight neutron spectroscopy --- the data show that the spectra indeed break four-fold symmetry with two distinguishable types of $\mathbf{q}$-vectors, both of which are gapped (see Supplementary Sec.~\ref{sec:SMtof}).

At 60~K in the ``spin-cholesteric'' phase [Fig.~\ref{CMP}(b2)] the quasi-elastic signal for $E$~<~1~meV is greatly enhanced at the suppressed wavevectors and becomes broad both in momentum and energy transfer due to the melting of long-range order in this direction. This quasi-elastic signal overshadows the gap, which was resolved at 1.5~K. The dispersion of excitations is not changed significantly, suggesting that the spectra consist of two contributions: the well-defined branches inherited from phase I and the quasi-elastic one. Above the N\'eel temperature [Fig.~\ref{CMP}(b3)], only the latter was observed, as the former vanishes once long-range order breaks down. The high-temperature short-range correlations are not suppressed with high field and can be observed as a diffuse signal, broad both in energy and momentum (detailed analysis in Supplementary Sec.~\ref{sec:SMfluct}).

In the vast majority of known compounds, the emergence of skyrmion phases stems from anti\-symmetric exchange (ASE) interactions, which are only allowed in a locally noncentrosymmetric environment. Systems combining such unconventional spin textures with spatial inversion are scarce. In the absence of ASE interactions, stabilization of topological magnetic structures may involve frustration, anisotropic exchanges, and long-range interactions mediated by itinerant electrons~\cite{Mostovoy_2005, Mostovoy_2005_2, Hayami_2021, Wang_2021, Okumura_2022, Hayami_2022}. Very recently, several new centrosymmetric systems with magnetic topological order were discovered, among them rare-earth based crystals of Gd$_3$RuAl$_{12}$~\cite{Hirschberger_2019, Hirschberger_2021}, Gd$_2$PdSi$_3$~\cite{Kurumaji_2019, Hirschberger_2020}, GdRu$_2$Si$_2$~\cite{Khanh_2020,Khan2022,Wood2023} GdRu$_2$Ge$_2$~\cite{Yoshimochi_2024}, and a single transition-metal compound SrFeO$_3$ which exhibits not only a skyrmion-lattice phase but a unique three-dimensional lattice of topological singularities known as hedgehog and antihedgehog spin textures~\cite{Ishiwata_2020}. Charge order in inversion-symmetric \sfo\ shifts the bridging O atoms slightly off the midpoint of the Fe--O--Fe exchange pathway~\cite{Peets2021}, but the magnetic order is the two-dimensional analogue of that in cubic-perovskite SrFeO$_3$ with nearly identical incommensurability, so \sfo\ presumably belongs to the list of multi-$\mathbf{q}$ magnets not relying on ASE interactions. Perovskites and their derivatives often have octahedra tilts and rotations which reduce the crystal symmetry, but these are absent in SrFeO$_3$ and \sfo. The hidden charge order in \sfo~\cite{Peets2021} leads only to slightly different moments at the two iron sites, but both sites order. The short correlation length of the superstructure reflections perpendicular to the Fe bilayers implies that each (resolution-limited) magnetic domain averages over numerous structural domains.

\begin{figure}
	\center{\includegraphics[width=1\columnwidth]{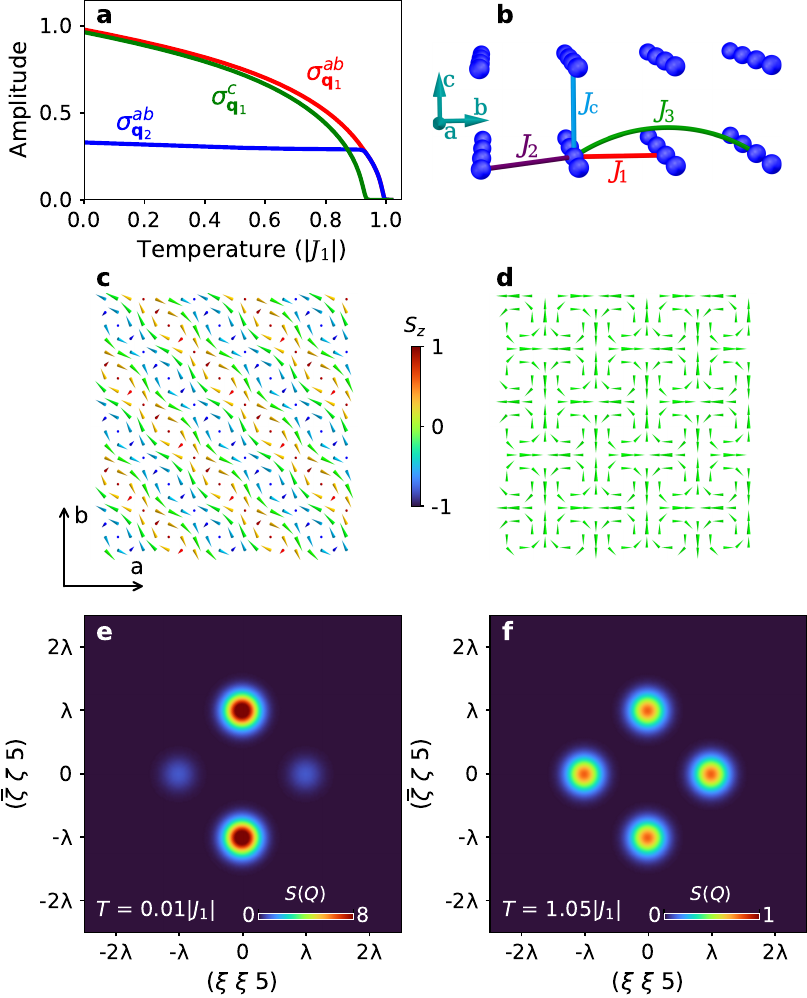}}
	\caption{Theoretical results. (a) Modulation intensities obtained from MFT calculations. (b) Exchange interactions within a bilayer of iron ions. (c), (d) Double-$\mathbf{q}$ ground state magnetic structure and high-temperature double-$\mathbf{q}$ state obtained from free energy minimization, respectively. (e), (f) Structure factor of magnetic textures in (c) and (d), respectively.}
	\label{four}
\end{figure}

Explaining the stability of double-$\mathbf{q}$ magnetic structures in the absence of ASE requires going beyond simple Heisenberg exchange interactions. Given that the experimental ground state in \sfo\ breaks $C_4$ rotational symmetry, we suggest competing anisotropic exchanges as the dominant source of frustration. The classical spin Hamiltonian with symmetric anisotropic exchanges can be written as follows:
\begin{align}
    E &= \frac{1}{2} \sum_{nm}{S^{\alpha}_{n} J^{\alpha\beta}_{\mathbf{x}_{n} - \mathbf{x}_{m}} S^{\beta}_{m}}, \nonumber\\
J^{\alpha\beta}_{\mathbf{r}} &= J_\mathbf{r} \delta^{\alpha\beta} + \mathrm{\Delta}_\mathbf{r}\bigl(\hat{r}^\alpha\hat{r}^\beta - \frac{1}{2}\delta^{\alpha\beta}\bigl),\ (\alpha,\beta = x,y), \nonumber\\
J^{zz}_{\mathbf{r}} &= J_\mathbf{r} + \delta_\mathbf{r}
    \label{eq:eq1}
\end{align}
where $\hat{\mathbf{r}} = \mathbf{r}/r$, and $\mathrm{\Delta}_\mathbf{r}$ and $\delta_\mathbf{r}$ are anisotropy parameters. The interactions considered in our analysis, drawn from Ref.~\citenum{Peets2014} and depicted in Fig.~\ref{four}(b), include intralayer exchanges up to third-nearest neighbours, with values of $J_1 = -7.2\text{ meV}$, $J_2 = 1.05\text{ meV}$, and $J_3 = 2.1\text{ meV}$, as well as the nearest intra-bilayer ferromagnetic exchange $J_c = -5.1\text{ meV}$; the non-zero anisotropy parameters $\mathrm{\Delta}_2 = -0.05|J_1|$ and $\delta_1 = 0.01|J_1|$ were used. Calculations of the phase diagram and magnetic ground state were carried out within mean field theory (MFT). As interbilayer interactions are reported to be small, a single square-lattice bilayer was considered. The energy in the mean-field approximation for a single bilayer can be written as
\begin{align}
    E_\text{MF} = -\sum_{n}{\mathbf{h}_n\cdot\mathbf{S}_n} -\frac{1}{2}\sum_{nm}{\langle S^{\alpha}_{n} \rangle J^{\alpha\beta}_{\mathbf{x}_{n} - \mathbf{x}_{m}} \langle S^{\beta}_{m} \rangle},
    \label{eq:eq2}
\end{align}
where $h_n^\alpha = -\sum_{m}{J^{\alpha\beta}_{\mathbf{x}_{n} - \mathbf{x}_{m}} \langle S^{\beta}_{m} \rangle}$ is the mean exchange field. This self-consistent equation allows determining the magnetic structure without {\itshape a priori} assumptions about it~\cite{Zhitomirsky_2022}. A detailed description of the calculations is in Supplementary Sec.~\ref{sec:numerics}.

As a result of our calculations, we describe the magnetic order by decomposing it into three main mutually orthogonal sinusoidal components: $\mathbf{\sigma}_{\mathbf{q}_1}^{ab}$ with wavevector $\mathbf{q}_1$ and in-plane spins $\mathbf{S}\perp\mathbf{q}_1$, an out-of-plane $\mathbf{\sigma}_{\mathbf{q}_1}^{c}$ modulation with the same vector $\mathbf{q}_1$, and in-plane $\mathbf{\sigma}_{\mathbf{q}_2}^{ab}$ with the orthogonal wave vector $\mathbf{q}_2$ and $\mathbf{S}\perp\mathbf{q}_2$. Figure~\ref{four}(a) shows the temperature dependence of the amplitudes of these modulations obtained from free-energy minimization in our MFT calculations, clearly evidencing the emergence of two different double-$\mathbf{q}$ phases. The ground state, the spin texture of which is shown in Figs.~\ref{four}(c,e), is formed by a helical $\mathbf{q}_1$-modulation and an in-plane sinusoidal $\mathbf{q}_2$-modulation. Our model predicts instability of the single-$\mathbf{q}$ helical state and unequal structure factors for the two types of ordering vectors, consistent with our low-temperature neutron diffraction data on \sfo. The out-of-plane spin component $\mathbf{\sigma}_{\mathbf{q}_1}^{c}$ decays more rapidly with temperature than does the in-plane $\mathbf{\sigma}_{\mathbf{q}_1}^{ab}$ until, above the critical temperature, two orthogonal in-plane modulations define the magnetic order. In this condition the structure factors of the two ordering vectors become equal, and the system transitions into a vortex-crystal state corresponding to the high temperature double-$\mathbf{q}$ phase III [Fig.~\ref{four}(d,f)]. The \sfo\ ``spin-cholesteric'' phase is not captured in our current simulations, since its properties are based crucially on spin fluctuations, which the MFT approximation cannot reproduce. Given that other key features of the model are consistent with experimental data, the anisotropy of symmetric exchange interactions presents a promising explanation for the magnetism observed in \sfo.

The low-temperature state with coexisting spiral and sinusoidal modulations was observed in the conducting non-collinear magnet, GdRu$_2$Si$_2$, with long-ranged interactions between spins~\cite{Khan2022}. The wave vectors of the two modulations are slightly different in that material, whereas in Sr$_3$Fe$_2$O$_7$ with nearly isotropic exchange interactions $q_1$ and $q_2$ are indistinguishable. This state was also found (at a nonzero magnetic field) in a model of a frustrated tetragonal magnet~\cite{Wang_2021}. The high-temperature planar $2\mathbf{q}$ state with equal amplitudes of the two modulations was observed in GdRu$_2$Si$_2$ at a nonzero magnetic field and was obtained using Landau theory near $T_{\rm N}$~\cite{Utesov2021}. These theoretical works, in addition to the $2\mathbf{q}$ states, predicted a variety of meron crystals with a nonzero net topological charge, which are not observed  in Sr$_3$Fe$_2$O$_7$ (at least, in zero magnetic field). Stability of meron crystals requires relatively large Kitaev-like interactions and single-ion anisotropy~\cite{Wang_2021} that would make magnetic susceptibility strongly anisotropic above $T_{\rm N}$, which is not the case in Sr$_3$Fe$_2$O$_7$. Across a helical spiral, a cycloidal spiral with spins in a vertical plane can also be unstable against the second modulation. Since cycloidal spirals induce an electric polarization through the inverse Dzyaloshinskii-Moriya mechanism, muliply-periodic states in such systems can be probed electrically.

Combining our theoretical and experimental results, it is evident that high-temperature low-field magnetic phase in \sfo\ is a planar sine+sine double-$\mathbf{q}$ vortex-crystal state. The ground state is a reentrant double-$\mathbf{q}$ state which marries helix and sine modulations and spontaneously breaks the $C_4$ rotational symmetry of the lattice in addition to chiral symmetry. The intermediate ``spin-cholesteric'' phase exhibits a completely new type of magnetic order, which is the magnetic analogue of a cholesteric. This phase is defined by a long-range ordered helix component at one propagation vector and strongly fluctuating short-range correlations in the orthogonal direction, arising from the suppressed sine component. The interplay between spiral order, the unique set of symmetry-breaking rules, and spin fluctuations in the ``spin-cholesteric'' phase suggests the potential emergence of novel properties. Our results establish \sfo\ as a highly promising candidate for studying multi-$\mathbf{q}$ order and exotic fluctuation-based magnetic states in centrosymmetric materials.

\vspace{0.8em}
{\footnotesize\noindent
{\textbf{Methods.}}

\noindent \textbf{Sample preparation.}
High-quality single crystals of Sr$_3$Fe$_2$O$_7$ were grown by the floating-zone technique\,\cite{Maljuk_2004,Peets2013}, then fully oxygenated under 5\,kbar of oxygen while gradually cooling from 450~$^\circ$C\,\cite{Peets2021}. A 4-g crystal was used for neutron measurements. \smallskip

\noindent \textbf{Neutron scattering experiments.}
Neutron diffraction was measured at the ZEBRA diffractometer at PSI, Villigen, Switzerland, using a 2D detector which allowed collecting data in three-dimensional reciprocal space in the vicinity of all four $\mathbf{Q} = (\pm\xi\pm\!\xi\,5)$ magnetic reflections in a single measurement sequence. The vertical focusing setup was utilized to improve the beam flux, resulting in elongation of peaks along the out-of-plane ($\overline{\mathbf{\xi}}\,\mathbf{\xi}\,0$) direction as a side effect. The diffraction patterns for the ($H\,K\,5$) plane in Fig.~\ref{CMP} were produced from rebinned data by integration along ($0\,0\,L$) for $4.88\leq L\leq5.12$. Peak widths were extracted similarly by integrating over two orthogonal directions in the close vicinity of every peak and by fitting a Gaussian to the third direction. The width errorbars represent standard deviations of fits. The integral intensity was obtained by integrating over all three directions.
Energy-resolved neutron scattering data were measured at the THALES triple-axis spectrometer and the IN5 time-of-flight spectrometer at the ILL, Grenoble, France.
\smallskip

\noindent \textbf{Magnetization measurements.}
The magnetization of a single crystal mounted to a plastic rod sample holder using GE varnish was measured as a function of temperature and magnetic field along [110] using the vibrating-sample magnetometry option in a Cryogenic Ltd.\ cryogen-free measurement system (CFMS). In addition to four-quadrant $M(H)$ loops and zero-field-cooled-warming and field-cooled-cooling measurements, field-trained data were collected after cooling to 40\,K in a 5\,T field, then to 2\,K in zero field. 
\smallskip

\noindent \textbf{Dilatometry measurements.}
Dilatometry measurements were performed using a tilted-plate capacitive dilatometer\cite{Rotter1998} with a sensitivity to relative length changes of $\sim$10$^{-7}$, using magnetic field sweep rates between 0.05 and 0.25~T/min. Measurements were made on single crystals for length changes parallel or perpendicular to the crystallographic [110] or [001] directions for magnetic fields oriented in both of these directions. This allows one to identify all magnetic transitions accompanied by lattice effects.
\smallskip

\noindent \textbf{Calculations.}
For a detailed description, see Sec.~\ref{sec:numerics} in the Supplementary Information.
\smallskip

\bibliography{sn-bibliography}

\smallskip
\noindent\textbf{Data availability.}
All relevant data are available from the authors upon reasonable request.

\noindent\textbf{Competing interests.}
The authors declare no competing interests.

\noindent\textbf{Acknowledgements.}
The authors are grateful for experimental assistance from the groups of M.\ Jansen and R.\ Kremer and for discussions with the group of Y.\ Motome. This project was funded by the German Research Foundation (DFG) through individual grants PE~3318/3-1, IN~209/7-1, and IN~209/9-1; through projects C01 and C03 of the Collaborative Research Center SFB~1143 (Project No.\ 247310070); through the W\"urzburg-Dresden Cluster of Excellence on Complexity and Topology in Quantum Materials\,---\,\textit{ct.qmat} (EXC~2147, Project No.\ 390858490); and through the Collaborative Research Center TRR~80 (Project No.\ 107745057). The authors acknowledge the Institut Laue-Langevin, Grenoble (France) for providing neutron beam time\,\cite{ILL_4-01-1589_IN5,ILL_4-01-1698}. Part of this work is based on experiments performed at the Swiss spallation neutron source SINQ, Paul Scherrer Institute, Villigen, Switzerland. \smallskip

\noindent{\textbf{Author~contributions.}}
D.S.I.\ conceived the idea and designed the experiments. Crystals were grown by D.C.P.\ under the guidance of B.K.  Magnetization measurements were performed by D.C.P.\ and S.G.\ under the guidance of B.K.\ and D.S.I., and the data were analyzed by N.D.A.\ and D.C.P. Dilatometry measurements were performed by J.G.\ and M.D., and the data were analyzed by J.G., M.D., and N.D.A.  Neutron diffraction measurements were performed by N.D.A., D.S.I., and O.Z., and the data were analyzed by N.D.A., N.S.P., and D.S.I. Inelastic neutron scattering was performed by N.S.P., J.O., A.I., Y.T., and D.S.I., and the data were analyzed by N.D.A., N.S.P., Y.T., and D.S.I. Theoretical calculations and analysis were performed by J.M.\ and M.M. The manuscript was written by N.D.A.\ with the assistance of D.C.P.\, D.S.I. and M.M., and all authors discussed the results and commented on the manuscript.
\smallskip

\vfill
\onecolumngrid\clearpage

%%%%SUPLEMENTARY
\renewcommand\thesection{S\arabic{section}}
\renewcommand\thefigure{S\arabic{figure}}
\renewcommand\thetable{S\arabic{table}}
\renewcommand\theequation{S\arabic{equation}}

\makeatletter
\renewcommand{\@oddfoot}{\hfill\bf\scriptsize\textsf{S\thepage}}
\renewcommand{\@evenfoot}{\bf\scriptsize\textsf{S\thepage}\hfill}
\renewcommand{\@oddhead}{N.\ D.\ Andriushin \textit{et~al.}\hfill\Large\textsf{\textcolor{NatureBlue}{SUPPLEMENTARY INFORMATION}}}
\renewcommand{\@evenhead}{H.~Jang \textit{et~al.}\Large\textsf{\textcolor{NatureBlue}{SUPPLEMENTARY INFORMATION}}\hfill}
\makeatother
\setcounter{page}{1}\setcounter{figure}{0}\setcounter{table}{0}\setcounter{equation}{0}

\onecolumngrid\normalsize

\begin{center}{\vspace*{0.1pt}\Large{Supplementary Information \smallskip\\\sl\textbf{``\hspace{1pt}Reentrant multiple-$\mathbf{q}$ magnetic order and a ``spin-cholesteric'' phase in Sr$_3$Fe$_2$O$_7$''}}}\end{center}\bigskip

\twocolumngrid

\section{Time-of-flight neutron scattering data} \label{sec:SMtof}
In an experimental setup with a triple-axis neutron spectrometer and vertical high-field magnet, the crystallographic direction orthogonal to the scattering plane is usually inaccessible. For this reason, in the case of THALES data, we are limited to the excitation spectrum above the suppressed $\mathbf{q}$-vectors orthogonal to the field and not able to cover the favored ones. However, for the low-temperature phase I corresponding to unequal double-$\mathbf{q}$ magnetic texture a time-of-flight INS experiment was performed on the IN5 instrument, where neutron scattering at both types of $\mathbf{q}$-vectors was measured.

Similar to the experiments described in the main text, the same \sfo\ sample was cooled down to base temperature in a field of 2.5~T $\mathbf{H} \parallel [\overline{1}10]$, resulting in a nearly-single-domain state. After cooling, the field was ramped down to zero and the measurements were performed at 1.8~K. From the elastic map covering all four magnetic satellites in vicinity of $\mathbf{Q} = (005)$ [Fig.~\ref{SMTOF}(a1)] it is revealed that both ordering vectors survived cooling in field and the double-$\mathbf{q}$ state features two types of unequally developed spin modulations.

Considering that the ground state is characterized by two unequal ordering vectors, it is expected that the excitations in the vicinity of these vectors would also differ. In the case of the \sfo\ phase I, although the $\mathbf{q}$-vectors exhibit significant differences in intensity, their incommensurability parameters are substantially identical. On account of this fact, the dispersions of spin wave excitations above the suppressed and favored $\mathbf{q}$-vectors are similar [Fig.~\ref{SMTOF}(b1,b2)]. The most evident dissimilarity lies in the divergence of two branches originating at incommensurate vectors, which is less pronounced for suppressed vectors (where the branches become resolved above $\sim$2~meV) compared to favored ones (where they are separated above 1.5~meV). The subtle details in spin excitations are clearer in constant energy cuts [Fig.~\ref{SMTOF}(a2-a4)]. The conical excitation bands cut at 2.5~meV form intensity rings with larger radii above the favored peaks. The patterns at 3.0~meV and 3.5~meV exhibit even more visibly broken four-fold symmetry, confirming the presence of two distinct types of ordering vectors in phase I.

\begin{figure}[t]
	\center{\includegraphics[width=1\columnwidth]{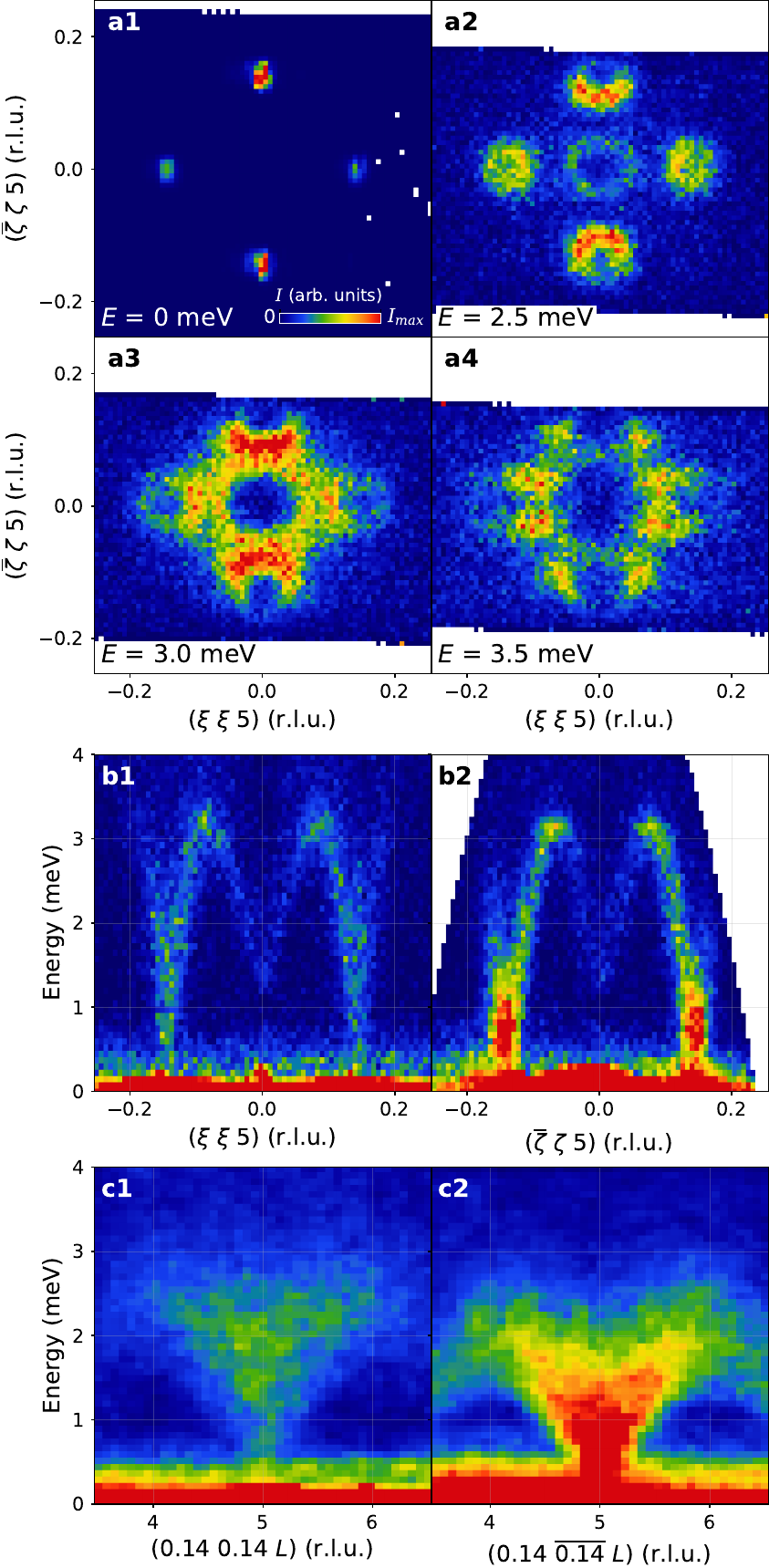}}
	\caption{Time-of-flight neutron scattering data on \sfo\ (IN5) measured at 1.8~K in zero field after cooling in 2.5~T. (a1-a4), Constant energy cuts in the ($HK$5) plane; the elastic colormap has a different colorscale. (b1,b2), Momentum-energy cuts for both suppressed (b1) and favored peaks (b2). (c1,c2), Momentum-energy cuts along the [001] direction: cutting through the suppressed (c1) and favored peaks (c2).}
	\label{SMTOF} \vspace{-20pt}
\end{figure}

Additionally, cuts along the [001] direction at suppressed and favored vectors are shown in Fig.~\ref{SMTOF}(c1,c2). In this direction, the excitation bands are determined by interactions between layers and bilayers and the antiferromagnetic coupling between subsequent layers results in forbidden magnetic Bragg peaks at even $L$ values. The excitations at suppressed peaks have higher bandwidth compared to their favored counterparts, which again demonstrates how \sfo\ has two types of inequivalent ordering vectors. The [001] spectra appear broader compared to cuts in the ($HK5$) plane due to the much larger $c$ lattice parameter.

The integration in close vicinity of magnetic Bragg peaks also highlights the inequivalence of suppressed and favored peaks (Fig.~\ref{1DTof}). The favored $\mathbf{q}$-vectors exhibit a gap of $\sim$0.7~meV, which is not resolved in the case of the suppressed peak (in the next section, it is shown using another experiment that excitations above suppressed peaks are also gapped). Conversely, the suppressed $\mathbf{q}$-vectors have a more prolonged and smeared tail, which corresponds to a steeper dispersion.

\begin{figure}
	\center{\includegraphics[width=1\columnwidth]{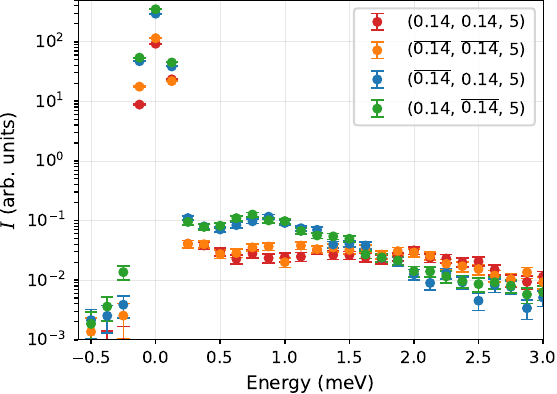}}
	\caption{Time-of-flight neutron scattering data on \sfo\ (IN5) measured at 1.8~K in zero field after cooling in 2.5~T. One-dimensional energy cuts for all four magnetic satellites.}
	\label{1DTof}
\end{figure}

\section{Neutron scattering in field} \label{sec:SMfield}

\subsection{Inelastic scattering.}
In the main text, the discussion regarding the excitations at low temperature was based on the energy-momentum map collected in high field [Fig.~\ref{CMP}(b1)]. In a field of ${\mu}_0H \approx\,3.5$~T and low temperatures, \sfo\ undergoes a phase transition observed in magnetization measurements. The magnetic excitations may potentially be influenced by the process of ramping up the field from zero to 10 T. Nevertheless, we found that the low-energy spectrum exhibits only minor changes in response to the magnetic field (see Fig.~\ref{FNF}). The most pronounced effect is a shift on the order of 0.1--0.2~meV, which is inconsequential for discussion of magnetic order. The important detail is the presence of an energy gap of 1~meV at the suppressed peak in both zero- and high-field cases. As was mentioned in the previous section, this gap at the suppressed ordering vector was not resolved in the time-of-flight INS experiment (Fig.~\ref{1DTof}). Combining the data from these two experiments, it is evident that \sfo\ in phase I has gapped excitations at both types of $\mathbf{q}$-vectors.
\begin{figure}[h]\vspace{10pt}
	\includegraphics[width=\columnwidth]{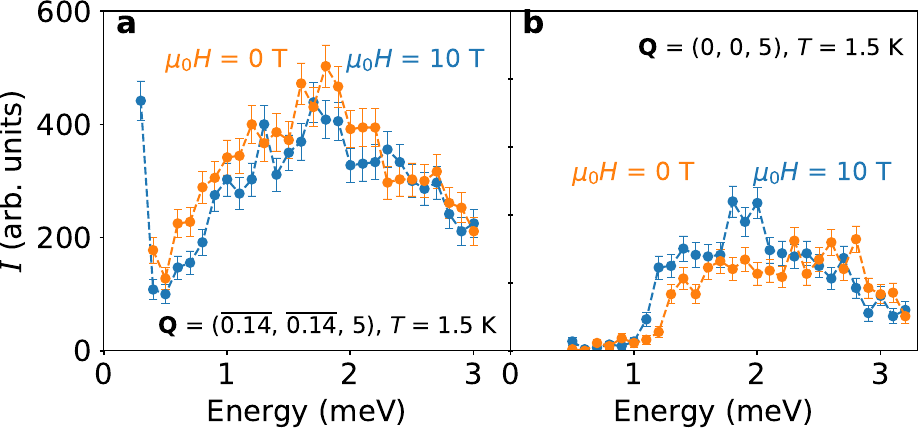}
	\caption{Comparison of inelastic neutron scattering (THALES) at $T$~=~1.5~K under zero and high field $\mathbf{H} \parallel [\overline{1}\,1\,0]$ (a), at the suppressed incommensurate magnetic peak $\mathbf{Q} = (-0.14, -0.14, 5)$ and (b), at the nuclear peak $\mathbf{Q} = (0, 0, 5)$.}
	\label{FNF}
\end{figure}
\subsection{Diffraction in field}
\begin{figure}
	\includegraphics[width=\columnwidth]{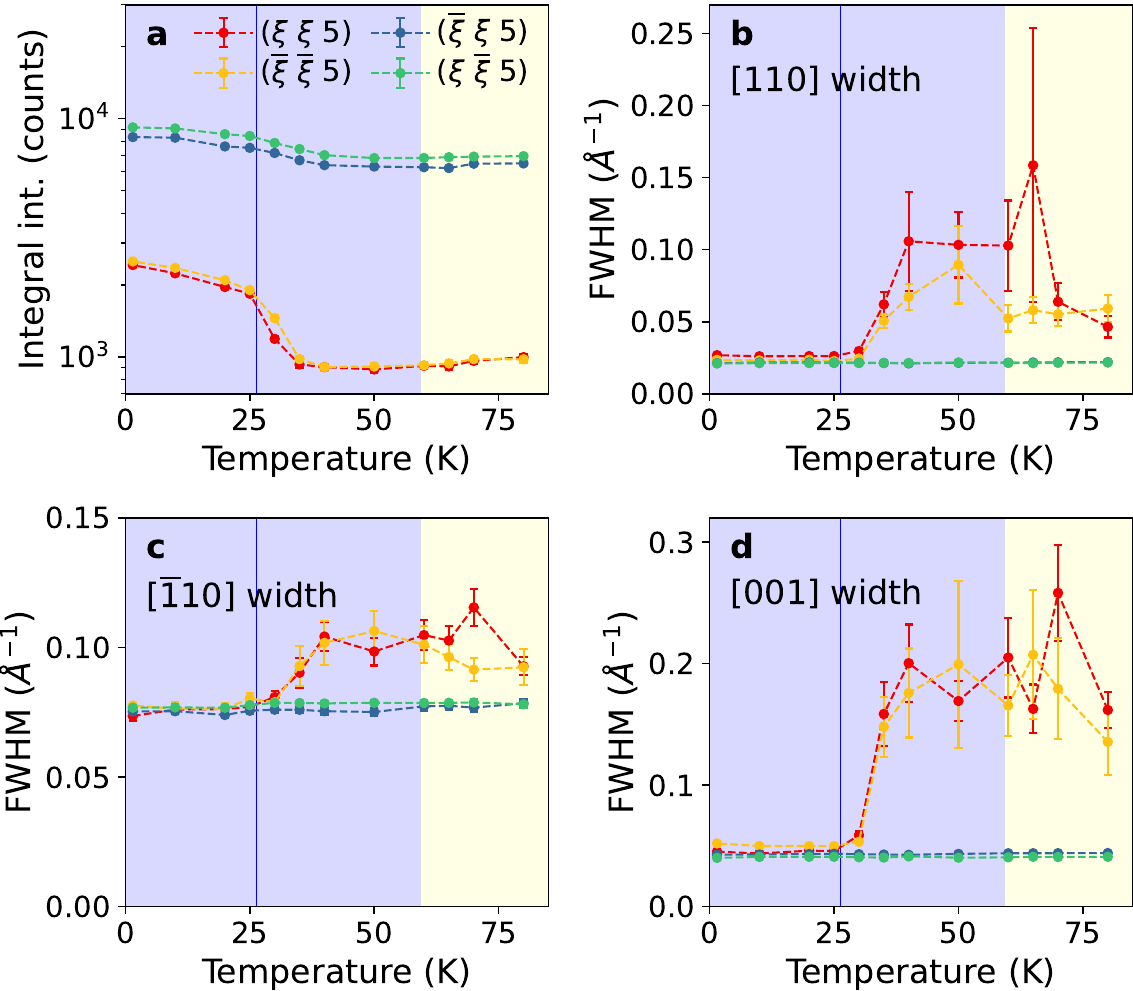}
	\caption{Single-crystal neutron diffraction data (ZEBRA). Data measured on cooling in field of 5.5~T: integral intensity (a) and peak widths along three orthogonal directions (b-d).}
	\label{fit_in_field}
\end{figure}

In the phase diagram [Fig.~\ref{PD}(c) in the main text] there are three high field phases denoted as phases IV$_\textrm{A}$, IV$_\textrm{B}$ and V. At a field of 5.5~T, the transitions among these phases were observed with magnetization measurements as the merging of ZFC\,---\,FC curves at 26~K and as a hump on the ZFC curve at 59~K. Using neutron diffraction, we can characterize the incommensurate magnetic satellites for the high field phases similarly to how we did it for the low field phases in the main text. After cooling in a field $\mathbf{H} \parallel [\overline{1}10]$ of 5.5~T, a sequence of measurements below 80~K was performed on the ZEBRA diffractometer (Fig.~\ref{fit_in_field}). The temperature regions corresponding to phases IV$_\textrm{A}$ and IV$_\textrm{B}$ (based on magnetization measurements) are shaded light blue, and the transition between them is denoted with a vertical line. The yellow background indicates phase V.

In phase V ($T$~>~59~K), the peaks orthogonal to the field are suppressed and broad in all momentum directions, while the peaks parallel to the field are resolution limited. This resembles phase II in low field, where similar broad suppressed peaks were observed. As temperature decreases, \sfo\ undergoes a transition into phase IV$_\textrm{B}$ with no noticeable changes in intensity or width of the incommensurate peaks. Similar to the magnetization measurements, it is possible that the transition V\,---\,IV$_\textrm{B}$ could be detectable upon warming in field after ZFC. At $T\approx25$~K the intensity of all four peaks increases and the suppressed peaks become sharp again. Similarly to phase I in low field, the intensity of the two types of peaks in phase IV$_\textrm{A}$ differs by a factor of $\sim\!$3--4.

\subsection{Field dependencies.}
The field dependencies of magnetization in \sfo\ can be used for observation of domain selection and a number of phase transitions. Naturally, these features manifest themselves in elastic neutron scattering as well. On the THALES triple-axis spectrometer, we measured elastic neutron scattering in the vicinity of the $(-0.14, -0.14, 3)$ peak, which is orthogonal to the field $\mathbf{H} \parallel [\overline{1}\,1\,0]$. At every field value, the ($H\,H\,3$) scan was measured and normalized against a beam monitor [the scans are shown in Fig.~\ref{fieldTHA}(e)]. For extraction of the amplitude $A$ of a peak, the scans were fitted with a Gaussian profile:
\begin{equation}
    I(H) = A e^{-(H - H_0)^2/\sigma^2} + I_0.
    \label{gauss}
\end{equation}
The magnetic domains in \sfo\ freeze below 25~K, and this can be seen on the comparison of ZFC and FC field dependencies of suppressed peak intensity at 1.5~K [Fig.~\ref{fieldTHA}(a,b)]. Cooling in field allows domain selection at higher temperature and reaches the single domain state. In contrast, a field of 10~T applied after ZFC leads to only a partial domain selection at 1.5~K with the intensity of the suppressed peak 3 times larger than under the same conditions but after FC. The phase transition at 4~T is also visible for both directions of the field sweep in agreement with magnetization measurements. In phase II the domains can be selected effectively in a field of 3~T [Fig.~\ref{fieldTHA}(c,d)].
\begin{figure}
	\includegraphics[width=\columnwidth]{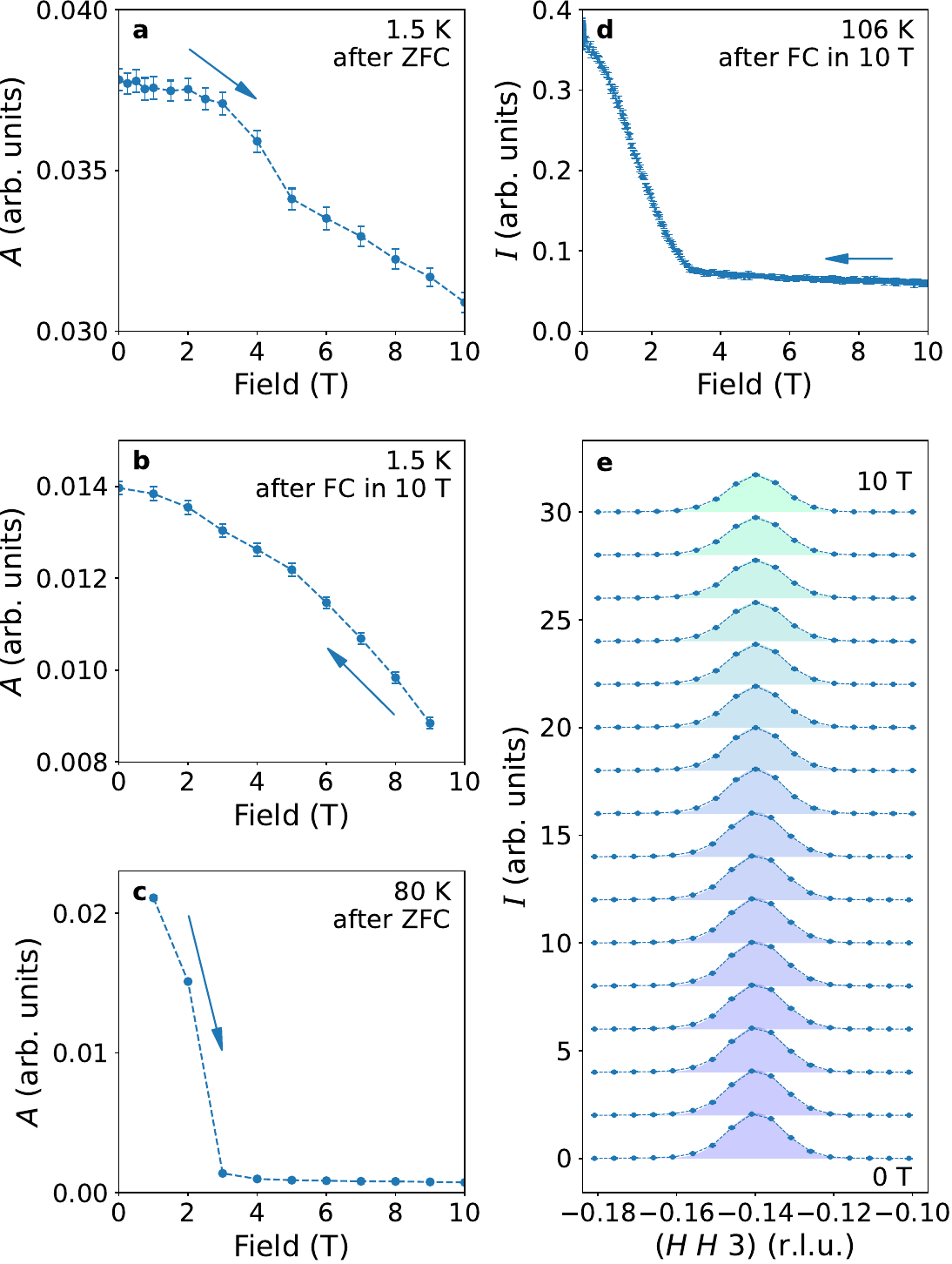}
	\caption{Field dependencies of neutron intensity [scattering vector $\mathbf{Q} = (-0.14, -0.14, 3)$ orthogonal to the field] measured at THALES. (a, b, c), Amplitudes of Gaussian peaks obtained with Eq.~\ref{gauss} fit, errorbars represent the estimated standard error for the best-fit value. (d), Intensity at the peak center as a function of magnetic field. (e), ($HH$3) scans at different external field values corresponding to panel (a). The filled areas are Gaussian fits, a constant offset is applied for clarity.}
	\label{fieldTHA}
\end{figure}

\section{High temperature fluctuations} \label{sec:SMfluct}
The fluctuations occurring near the critical temperature in \sfo\ can be effectively characterized by the broadening of magnetic reflection peaks. Here, we measured the temperature evolution of quasi-elastic intensity at the ordering vector $\mathbf{Q} = (-0.13, -0.13, 5)$ (Fig.~\ref{gamma}). As expected for a frustrated magnet, fluctuations play a major role in the vicinity of the ordering temperature (phases II and III) and persist far above it. The observed signal exhibits a broad Lorentzian-like profile, with a Gaussian component on top of it, originating from the ordered magnetic moments below $T$~117~K and from incoherent scattering at higher temperatures. The fluctuations' characteristic correlation timescale was extracted by fitting the following equation:
\begin{gather}
    I(E) = I_{\mathrm{Gau}}(E, E_0 = 0) + I_{\mathrm{Lor}}(E, E_0 = 0) + I_0, \nonumber\\
    I_{\mathrm{Lor}}(E, E_0) = \frac{A E}{1 - \mathrm{exp}(-E/k_{\mathrm{B}}T)} \times \nonumber\\
    \times\Big( \frac{\Gamma}{(E - E_0)^2 + \Gamma^2} +  \frac{\Gamma}{(E + E_0)^2 + \Gamma^2}\Big),\label{eq:lorentz}
\end{gather}
where the first term $I_{\mathrm{Gau}}$ is a Gaussian contribution, the second one $I_{\mathrm{Lor}}(E, E_0)$ is a general equation for Lorentzian with full-width-half-maximum $\Gamma$ and amplitude $A$, and $I_0$ is a constant background. The Lorentzian peak has two terms, corresponding to energy gain and loss processes, which however only differ if $E_0$ is nonzero.

\begin{figure}\vspace{10pt}
	\includegraphics[width=\columnwidth]{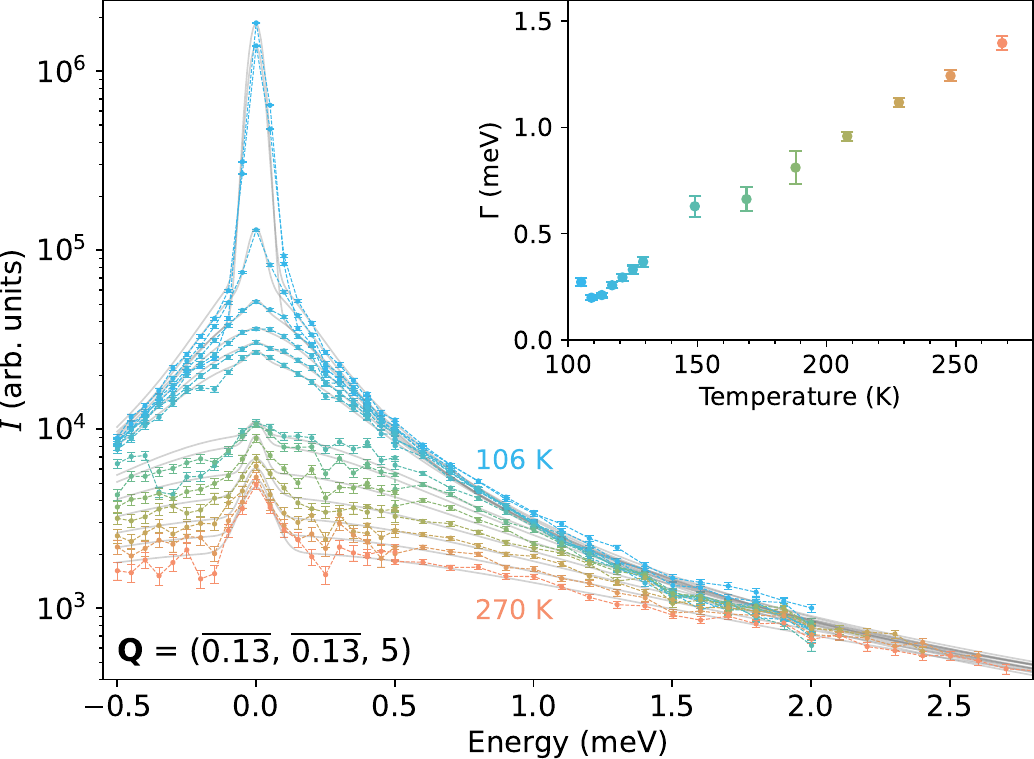}
	\caption{Quasi-elastic neutron scattering at different temperatures. Inset shows the Lorentzian FWHM $\Gamma$ representing the characteristic timescale of fluctuations obtained by fitting with Eq.~\ref{eq:lorentz} (grey curves).}
	\label{gamma}
\end{figure}

\begin{figure}
	\includegraphics[width=\columnwidth]{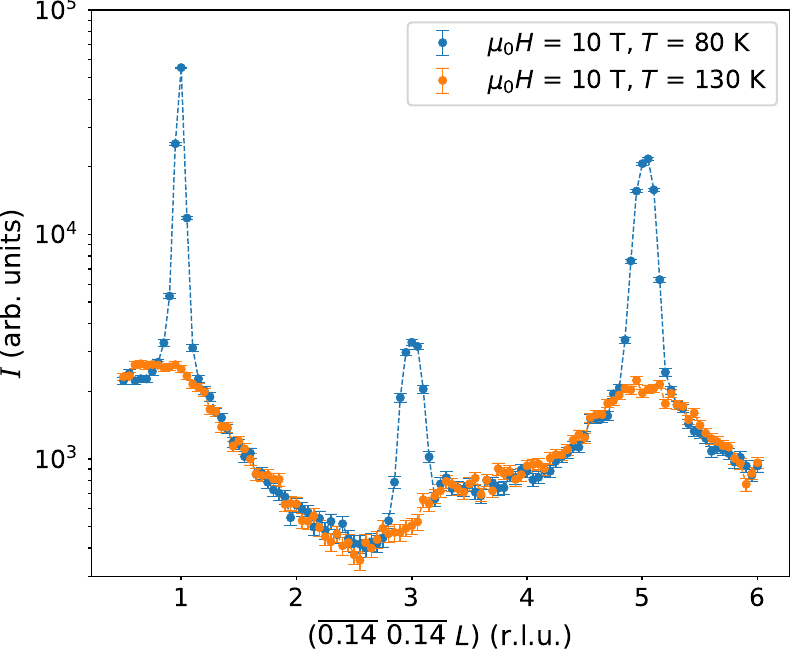}
	\caption{Elastic scattering intensity for the suppressed magnetic peaks (THALES).}
	\label{diffuse00L}
\end{figure}

The high-temperature short-range order is manifested in momentum space in a form of diffuse signal. The smearing of the magnetic peaks discussed in the main text [Fig.~\ref{I_vs_T}(c-e)] is a natural indicator of how short-range correlations take the place of long-range order. The elastic data along the [001] direction measured in high field demonstrate that these correlations are preserved even in 10~T (Fig.~\ref{diffuse00L}). Below the ordering temperature, the magnetic peaks have a well-defined Gaussian profile, which is located on top of a strong diffuse background. Heating up shows that the diffuse signal keeps its form in the paramagnetic phase, despite being enhanced with a contribution from the smeared Bragg peaks. This leads us to conclude that the short-range correlations in \sfo\ are robust against external field and persist throughout a broad temperature region.

\section{Relaxation of the magnetic order} \label{sec:SMrelax}
The magnetic order in \sfo\ is accompanied at higher temperatures by strong fluctuations of domain walls, which leads to domain relaxation toward the magnetic multidomain state. In Fig.~\ref{relax} we show time-dependent measurements of the elastic neutron scattering intensity at the suppressed ordering vector at $T$~=~100~K. An external field $\mathbf{H} \parallel [\overline{1}10]$ of 4~T was applied and then reduced to zero after the intensity saturated. The data show how the magnetic subsystem returns to its equilibrium state after the field reached zero ($t=0$). The data are well described by a stretched exponential,
\begin{equation}
    M(t) = M_\infty + M_0 e^{-\sqrt{t/\tau}}.
    \label{eq:relax}
\end{equation}
The characteristic domain relaxation time obtained from this fit is $\tau$~=~4397~s at 100~K, which is sufficiently slow to be observable in a neutron diffraction measurement through the time dependence of the magnetic Bragg peak intensity. This slow relaxation likely underlies the discrepancies in the temperature dependence in Fig.~\ref{I_vs_T}(a) around 90~K, where the heating or cooling rate can have a significant influence.
\begin{figure}\vspace{-6pt}
	\includegraphics[width=\columnwidth]{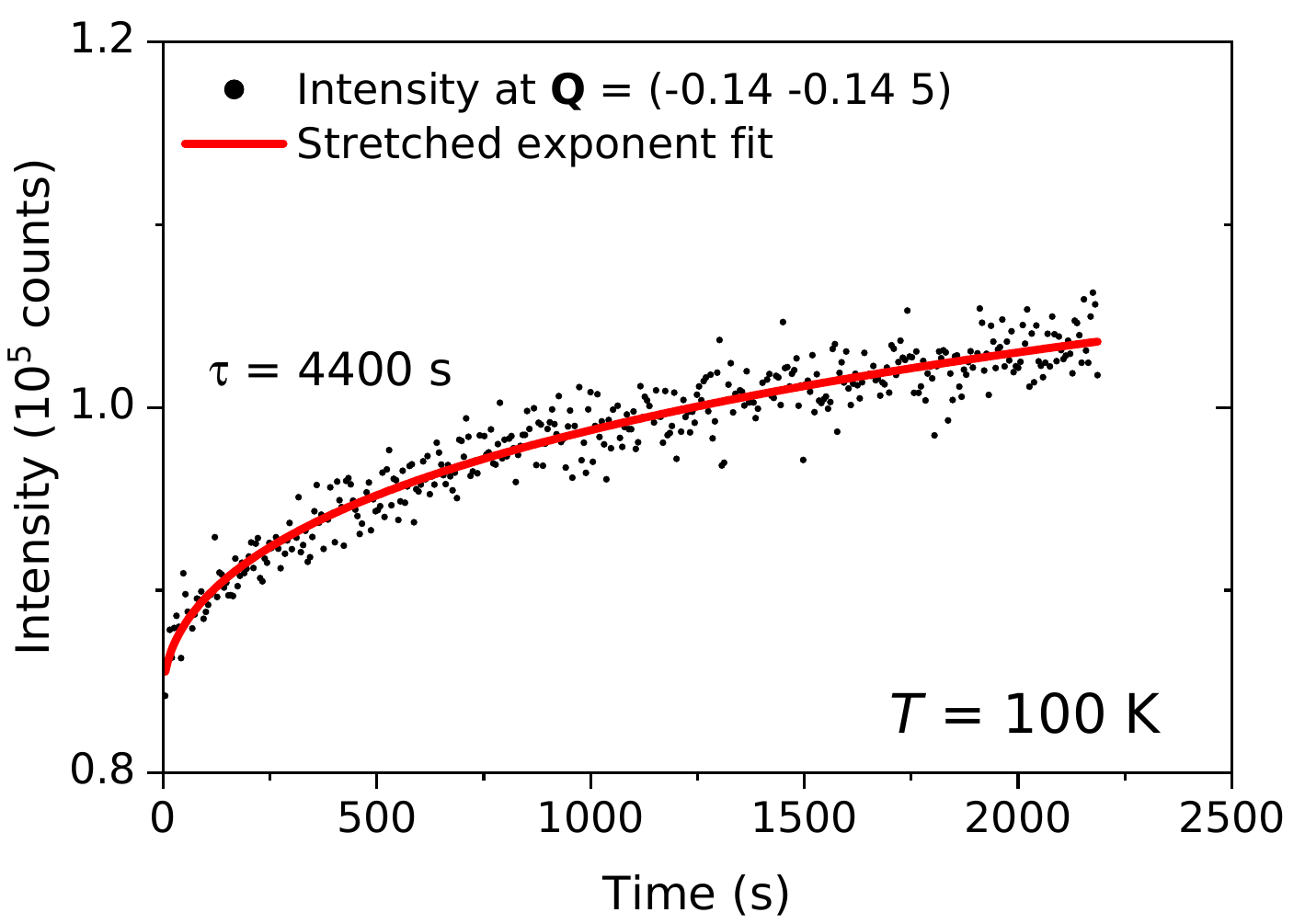}
	\caption{Time dependence of elastic neutron scattering intensity at $\mathbf{Q} = (-0.14,~-0.14,~5)$ after the external field is switched off. The red curve is a stretched-exponential fit to Eq.~\ref{eq:relax}.}
	\label{relax}
\end{figure}

\section{Magnetization} \label{sec:SMmagnetiz}

\begin{figure}
	\includegraphics[width=\columnwidth]{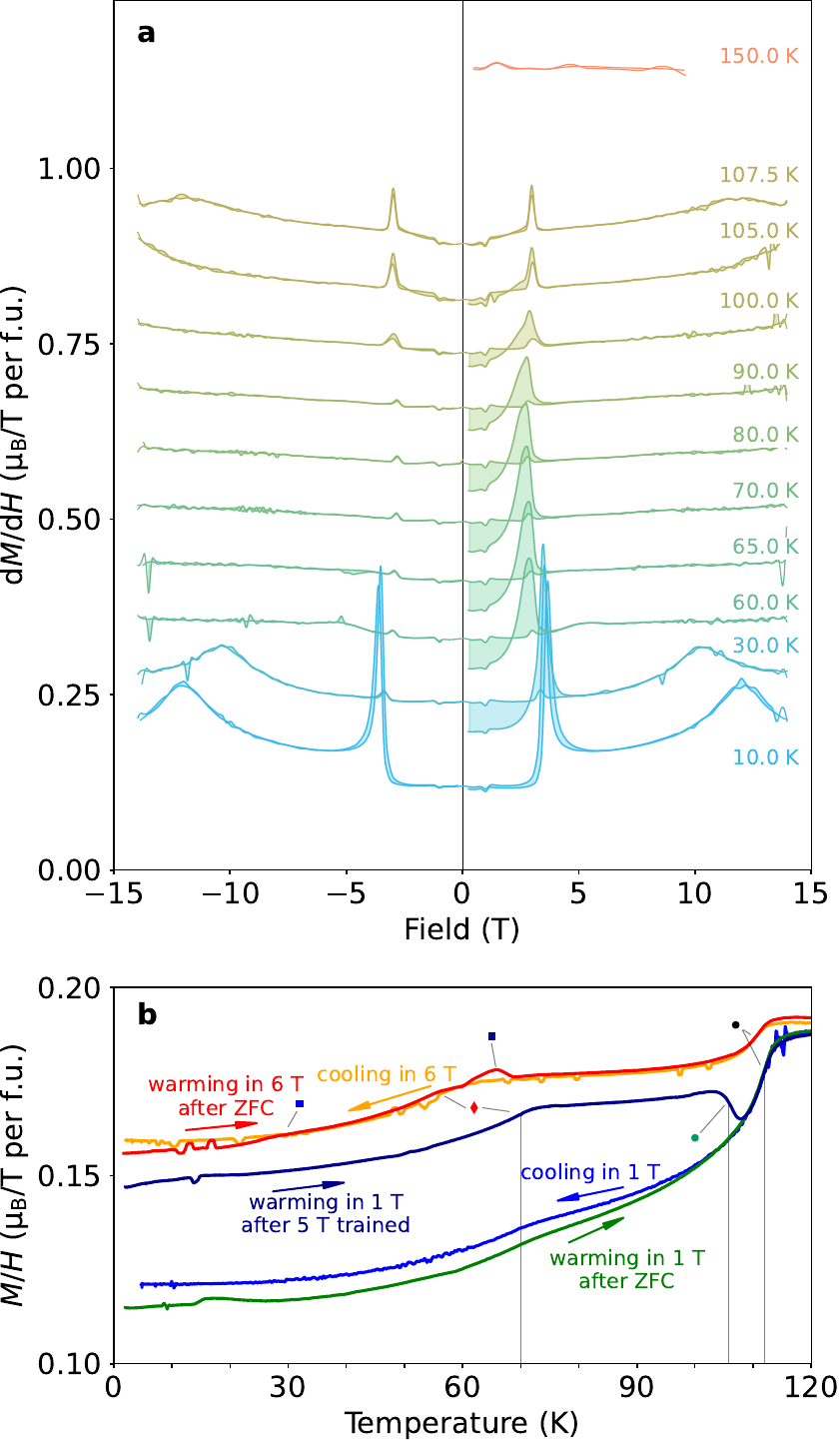}
	\caption{Magnetization measurements in \sfo\ for the $[\overline{1}\,1\,0]$ field direction. (a), The derivative of magnetization with respect to the field at different temperatures. Each curve includes a four-quadrant $M$-$H$ loop; the difference between field ramping up and down is shaded. A constant offset is applied for visual clarity. (b), Temperature dependence of magnetization. The symbols denoting features in the data correspond to those in the phase diagram in the main text [Fig.~\ref{PD}(c)].}
	\label{magnetization}
\end{figure}

The magnetic phase diagram of \sfo\ in Fig.~\ref{PD} in the main text was constructed using magnetization and dilatometry measurements. The magnetization features corresponding to the transitions, and how they are decoded in the phase diagram plot, are shown in Fig.~\ref{magnetization}(b). For the discussion of magnetic order in the low-temperature phase the magnetization data is relevant as it demonstrates how field training can drive \sfo\ into a single-domain state. For instance, it is notable how the transition from phase II to phase III (green circle) can be totally overlooked if measurements are done with only ZFC and FC prehistory. The transition between phase IV$_a$ and phase IV$_b$ is extracted as a merging point of two curves with ZFC and FC protocols. The effect of field training on magnetization is also visible in Fig.~\ref{magnetization}(a). Since each $M(H)$ curve was measured after ZFC, the irreversible effect of domain selection is observed and shown with shading representing the difference between field ramping up and down. The magnetization measurements are discussed in more detail in Ref.~\citenum{PhaseDiag}.

\section{Numerical Simulations}
\label{sec:numerics}

\subsection{Anisotropic Spin Interactions}
\label{sec:interactions}

The strong ferromagnetic interlayer coupling within the bilayer aligns spins in neighboring layers parallel to each other, which allows us to consider an effective single-layer spin model on a square lattice. The 4/$mmm$ site symmetry of this model forbids antisymmetric Dzyaloshinskii-Moriya interactions between any two spins, as well as the $zx$ and $zy$ symmetric anisotropic exchange interactions:
\begin{equation}
J_{\mathbf{r}}^{\alpha\beta} = 
\left[
\begin{array}{ccc}
\hspace*{-0.5ex}J_{\mathbf{r}} + a_{\mathbf{r}}(\hat{r}^x\hat{r}^x -\hat{r}^y\hat{r}^y)  & b_{\mathbf{r}}\hat{r}^x\hat{r}^y & 0\\
b_{\mathbf{r}}\hat{r}^x\hat{r}^y & \hspace{-1.5ex}J_{\mathbf{r}} - a_{\mathbf{r}}(\hat{r}^x\hat{r}^x -\hat{r}^y\hat{r}^y)\hspace*{-1.5ex} & 0\\
0 & 0 & J_{\mathbf{r}} + \delta_{\mathbf{r}}\hspace*{-0.5ex}
\end{array}
\right],
\label{eq:Jr1}
\end{equation}
where $\mathbf{r}$ is the radius vector separating two sites, with the corresponding unit vector $\hat{\mathbf{r}} = \frac{\mathbf{r}}{|\mathbf{r}|}$. In the model with interactions up to third-nearest neighbor, either $a_{\mathbf{r}}$ or $b_{\mathbf{r}}$ is 0 and Eq.~(\ref{eq:Jr1}) can be written in a simpler form (see Eq.~\ref{eq:eq1} in the main text).

The modulation wave vector $\mathbf{q}$ is determined by the smallest eigenvalue of the Fourier transform of the exchange constants matrix:
\begin{equation}
J_{\mathbf{q}}^{\alpha\beta} = \sum_{\mathbf{r}} J_{\mathbf{r}}^{\alpha\beta}e^{-i \mathbf{q}\cdot\mathbf{r}}.
\end{equation}
While the direction and length of $\mathbf{q}$  are largely governed by isotropic Heisenberg interactions, the relatively weak exchange anisotropies control the spin polarization of magnetic modulations and play an important role in stabilization of multiply-periodic states.

\subsection{Unrestricted Mean Field Calculations}
\label{sec:meanfield}

The average spin vector in the mean field approximation (see Eq.~\ref{eq:eq2} in the main text) is given by
\begin{equation}
\langle \mathbf{S}_{n} \rangle = S \frac{\mathbf{h}_{n}}{h_{n}} L(\beta S h_{n}),
\label{eq:selfconsistency}
\end{equation}
where $S$ is the spin length, $\mathbf{h}_{n}$ is the mean exchange field, $\beta = \frac{1}{k_{\rm B}T}$, $k_{\rm B}$ is the Boltzmann constant, and $L(x) = \coth(x) - \frac{1}{x}$ is the Langevin function. The self-consistency condition, $h^\alpha_{n} = - \sum\limits_{m} J_{\mathbf{x}_n -\mathbf{x}_m}^{\alpha\beta} \langle S_{m}^\beta \rangle$, is reached by iterations at all lattice sites of the bilayer with $21\times21\time2$ spins. Out of the spin configurations satisfying the self-consistency condition, we chose the one that minimizes the free energy 
\begin{equation}
F_{\rm MF} = - k_{\rm B}  T \sum_{n}
\ln
\left(
\frac{\sinh( \beta S h_n)}{ \beta S h_n}
\right) 
-
\frac{1}{2} 
\sum_{nm}
\left\langle S_{n}^\alpha \right\rangle
J_{\mathbf{x}_n -\mathbf{x}_m}^{\alpha\beta} 
\left \langle S_{m}^\beta \right \rangle.
\label{eq:FMF}
\end{equation}

\subsection{Low-temperature $\mathbf{2q}$ state}
\label{sec:instability}

The appearance of the $2\mathbf{q}$ state with unequal amplitudes of the modulations at low temperatures can be understood as follows. Consider three sinusoidal modulations, 
$\mathbf{S}_{n} = \mathbf{e}_{\mathbf{q}\lambda}\sin (\mathbf{q}\cdot \mathbf{x}_n)$, where $\mathbf{e}_{\mathbf{q}\lambda}$ are eigenvectors of the matrix 
$J_{\mathbf{q}}^{\alpha\beta}$: 
\begin{equation}
J_{\mathbf{q}}^{\alpha\beta} e_{\mathbf{q}\lambda}^\beta = 
J_{\mathbf{q}\lambda} e_{\mathbf{q}\lambda}^\alpha.
\end{equation}
The index $\lambda = 1,2,3$ describes the spin polarization of the modulation: 
\begin{equation}
\mathbf{e}_{\mathbf{q}1} = \hat{\mathbf{q}},\quad 
\mathbf{e}_{\mathbf{q}2} = \hat{\mathbf{z}}\quad 
\mbox{and}\quad
\mathbf{e}_{\mathbf{q}3} = \hat{\mathbf{q}} \times \hat{\mathbf{z}},
\end{equation} 
i.e. spins are polarized, respectively, along the wave vector, along the $c$ axis and in the $ab$-plane perpendicularly to the wave vector. The helical spiral with the wave vector $\mathbf{q}_1$ is unstable, if the $\lambda = 1$ modulation has the highest energy and the $\lambda = 3$ modulation has the lowest energy: ${\cal E}_{\mathbf{q}1} > {\cal E}_{\mathbf{q}2} > {\cal E}_{\mathbf{q}3}$, where ${\cal E}_{\mathbf{q}\lambda} =  \frac{S^2}{2}J_{\mathbf{q}\lambda}$ is the energy of the modulated state per lattice site. Equivalently, the energy of the helical spiral, $\frac{1}{2}({\cal E}_{\mathbf{q}2} + {\cal E}_{\mathbf{q}3})$, is lower than the energies of two cycloidal spirals: 
$\frac{1}{2}({\cal E}_{\mathbf{q}1} + {\cal E}_{\mathbf{q}3})$, for the spiral with spins in the $ab$-plane, and  $\frac{1}{2}({\cal E}_{\mathbf{q}1} + {\cal E}_{\mathbf{q}2})$, for the spiral with spins in the vertical plane. 

When an additional sinusoidal modulation with the transverse wave vector $\mathbf{q}_2$ and a small amplitude $\varepsilon$ appears, the amplitude of the helical modulation is reduced by approximately  $(1 - \frac{\varepsilon^2}{2})$ to preserve the spin length, and the energy of the $2\mathbf{q}$ state, $\varepsilon^2 {\cal E}_{\mathbf{q}_2 3} + \frac{(1 - \varepsilon^2)}{2}({\cal E}_{\mathbf{q}_1 2} + {\cal E}_{\mathbf{q}_1 3})$, is lower than the energy of the helical spiral by $\frac{\varepsilon^2}{2}({\cal E}_{\mathbf{q}_1 2}-{\cal E}_{\mathbf{q}_1 3})$, where we used $ {\cal E}_{\mathbf{q}_2 3} =  {\cal E}_{\mathbf{q}_1 3}$ following from $4_z$ symmetry. We note that the instability also occurs for ${\cal E}_{\mathbf{q}3} > {\cal E}_{\mathbf{q}2} > {\cal E}_{\mathbf{q}1}$, i.e. a cycloidal spiral with spins in a vertical plane is unstable against a second modulation provided the $c$ axis is the intermediate magnetic axis.

\subsection{High-temperature $\mathbf{2q}$ state}
\label{sec:landau}

The emergence of the $2\mathbf{q}$ state with equal amplitudes of two modulations at elevated temperatures can be understood using the Landau expansion of the free energy Eq.~\ref{eq:FMF} near the N\'eel temperature, $T_{\rm N}$:
\begin{equation}
k_{\rm B}  T_{\rm N} = - \frac{S^2}{3}   J_{\mathbf{q}_1 3}.
\end{equation} 

The general form of the ordered state just below $T_{\rm N}$ is,
\begin{equation}
\mathbf{H}_n = \beta S\mathbf{h}_{n} =  \mathrm{\Delta}_{\perp1}\mathbf{e}_{\mathbf{q}_{1}3}
\cos (\mathbf{q}_1\cdot \mathbf{x}_n) +   
\mathrm{\Delta}_{\perp2}\mathbf{e}_{\mathbf{q}_{2}3}
\cos (\mathbf{q}_2\cdot \mathbf{x}_n). 
\label{eq:Hn1}
\end{equation}
The phases of the two modulations are arbitrary and can be eliminated by shifting the spin configurations along $\mathbf{q}_1$ and $\mathbf{q}_2$. Substituting Eq.~\ref{eq:Hn1} into Eq.~\ref{eq:FMF}, we obtain an expression for the free energy per site in terms of the order parameters, $\mathrm{\Delta}_{\perp1}$ and $\mathrm{\Delta}_{\perp2}$:
\begin{multline}
f_{\rm MF} = \frac{F_{\rm MF}}{N} = \frac{k_{\rm B}  T}{12}\Biggl[ - \tau\left(\mathrm{\Delta}_{\perp1}^2+\mathrm{\Delta}_{\perp2}^2 \right) \\
+\frac{1}{40}\left(\left(\mathrm{\Delta}_{\perp1}^2+\mathrm{\Delta}_{\perp2}^2 \right)^2-\frac{2}{3}\mathrm{\Delta}_{\perp1}^2 \mathrm{\Delta}_{\perp2}^2\right)\Biggr],
\label{eq:fMF1}
\end{multline}
where $\tau = \frac{T_{\rm N} - T}{T_{\rm N}}$. The  negative  fourth-order anisotropy term,  $\mathrm{\Delta}_{\perp1}^2 \mathrm{\Delta}_{\perp2}^2$, makes the free energy of the $2\mathbf{q}$ state with equal amplitudes of the two modulations, $\mathrm{\Delta}_{\perp 1} = \pm \mathrm{\Delta}_{\perp2}$, lower than that of the single-${\mathbf q}$ modulation, e.g. $\mathrm{\Delta}_{\perp1} = \mathrm{\Delta}_{\perp},\mathrm{\Delta}_{\perp2} = 0$, by $\frac{1}{6}k_{\rm B} T_N \tau^2$. 

Usually, when an ordered state appears, it tends to suppress other competing orders. The state with $\mathrm{\Delta}_{\perp 1} = \pm \mathrm{\Delta}_{\perp2}$ is stable, because (i) spins in the two ordered states are orthogonal to each other and (ii) spins are periodically modulated rather than spatially uniform.  Fourth-order terms in the expansion of free energy give rise to higher harmonics, $\cos (3 \mathbf{q}_1\cdot \mathbf{x}_n)$ and  $\cos (3 \mathbf{q}_2\cdot \mathbf{x}_n)$, but their contribution to the free energy is $\propto \tau^3$ and can be neglected for small $\tau$.

As temperature decreases, the state with 
$\mathrm{\Delta}_{\perp 1} = \pm \mathrm{\Delta}_{\perp2} = \frac{\mathrm{\Delta}_{\perp}}{\sqrt{2}}$ 
becomes unstable towards an additional  modulation along $\mathbf{e}_{\mathbf{q}_{1}2} = \mathbf{e}_{\mathbf{q}_{2}2} = \hat{\mathbf{z}}$. 
The Landau expansion of the free energy of the $2\mathbf{q}$ state including the two out-of-plane modulations, 
\begin{align}
\mathbf{H}_n =  
&\frac{\mathrm{\Delta}_{\perp}}{\sqrt{2}}
\mathbf{e}_{\mathbf{q}_{1}3}
\cos (\mathbf{q}_1\cdot \mathbf{x}_n) +   
\mathrm{\Delta}_{z1}
\mathbf{e}_{\mathbf{q}_{1}2}
\sin (\mathbf{q}_1\cdot \mathbf{x}_n)
\nonumber\\
+
&\frac{\mathrm{\Delta}_{\perp}}{\sqrt{2}}
\mathbf{e}_{\mathbf{q}_{2}3}
\cos (\mathbf{q}_2\cdot \mathbf{x}_n)+
\mathrm{\Delta}_{z2}
\mathbf{e}_{\mathbf{q}_{2}2}
\sin (\mathbf{q}_2\cdot \mathbf{x}_n)
, 
\label{eq:Hn2}
\end{align}
is
\begin{align}
f_{\rm MF} = \frac{k_{\rm B}  T}{12}
&\left[ 
\left(
- \tau \mathrm{\Delta}_{\perp}^2 
+ \frac{\mathrm{\Delta}_{\perp}^4}{48}
\right)
\right. \nonumber \\
&\left.
+\left(- \frac{\tau - \frac{\delta T}{T_{\rm N}}}
{1 -   \frac{\delta T}{T_{\rm N}}}
+  \frac{\mathrm{\Delta}_{\perp}^2}{40}
\right)
\left(\mathrm{\Delta}_{z1}^2+\Delta_{z2}^2 \right)
\right. \nonumber \\
&\left.
+ \frac{1}{40} 
\left(
\left(
\mathrm{\Delta}_{z1}^2 + \mathrm{\Delta}_{z2}^2
\right)^2
+ 2 \mathrm{\Delta}_{z1}^2 \mathrm{\Delta}_{z2}^2
\right)
\right],
\label{eq:fMF2}
\end{align}
where $\delta T = \frac{S^2}{3}\left(J_{\mathbf{q}_1 2} - J_{\mathbf{q}_1 3}\right)>0$. From Eq.~\ref{eq:fMF2} we find that the out-of-plane modulation appears at $T_z  = (1 - \tau_{z})T_{\rm N}$, where
\begin{equation}
\tau_{z} = \frac{5\frac{\delta T}{T_{\rm N}} }
{2  + 3\frac{\delta T}{T_{\rm N}}}.
\end{equation}
Since spins in the two out-of-plane modulations are parallel to each other, the coefficient in front of $\mathrm{\Delta}_{z1}^2 \mathrm{\Delta}_{z2}^2$ is positive, which implies that only one of these orders is present at $T < T_z$. This in turn makes  the in-plane modulations asymmetric: $\mathrm{\Delta}_{\perp1} \neq \mathrm{\Delta}_{\perp2}$ below $T_z$.
\vfill

\end{document}